\documentclass[aps,prb,twocolumn,showpacs,preprintnumbers,amsmath,amssymb]{revtex4-1}

\usepackage{graphicx}
\usepackage{dcolumn}
\usepackage{bm}

\usepackage[english]{babel}
\usepackage[utf8]{inputenc}
\usepackage{amssymb}
\usepackage{amsmath}
\usepackage{color}
\usepackage{hyperref}
\usepackage{bbm}
\usepackage{epsfig}

\usepackage{umoline}

\newcommand{\e}{\varepsilon}
\newcommand{\s}{\sigma}
\newcommand{\w}{\omega}
\newcommand{\G}{\Gamma}

\newcommand{\bk}{\bm k}
\newcommand{\sbar}{\bar\sigma}
\newcommand{\smin}{\! - \!}
\renewcommand{\Im}{{\rm Im}\,\,}

\newcommand{\de}{{\rm d}}
\newcommand{\dd}{\partial}

\renewcommand{\P}{\mathcal{P}}
\newcommand{\A}{\mathcal{A}}

\newcommand{\dk}{d^\dagger}

\newcommand{\ra}{\rangle}
\newcommand{\la}{\langle}
\newcommand{\up}{\uparrow}
\newcommand{\down}{\downarrow}

\newcommand{\GF}[1]{ {\la\!\la #1 \ra\!\ra } }

\newcommand{\beq}{ \begin{equation} } 
\newcommand{\eeq}{ \end{equation} }
\newcommand{\beqa}{\begin{eqnarray}}
\newcommand{\eeqa}{\end{eqnarray}}
\newcommand{\nn}{\nonumber}
\newcommand{\es}{& = &}

\newcommand{\fig}[1]{Fig.~\ref{#1}}
\newcommand{\eq}[1]{Eq.~(\ref{#1})}

\newcommand{\exch}{\Delta\varepsilon_{\rm exch}}

\newcommand{\pol}{\mathcal{P}}

\begin{document}

\title{Perfect spin polarization in T-shaped double quantum dots \\due to the spin-dependent Fano effect}

\author{Krzysztof P. W{\'o}jcik}
\email{kpwojcik@amu.edu.pl}
\author{Ireneusz Weymann}
\email{weymann@amu.edu.pl}
\affiliation{Faculty of Physics, Adam Mickiewicz University, Umultowska 85, 61-614 Pozna{\'n}, Poland}
\date{\today}

\begin{abstract}
We study the spin-resolved transport properties of T-shaped double quantum dots
coupled to ferromagnetic leads. Using the numerical renormalization group method,
we calculate the linear conductance and the spin polarization
of the current for various model parameters and at different temperatures.
We show that an effective exchange field due to the presence
of ferromagnets results in different conditions for Fano destructive interference
in each spin channel. This spin dependence of the Fano effect leads to perfect spin polarization,
the sign of which can be changed by tuning the dots' levels.
Large spin polarization occurs due to Coulomb correlations
in the dot, which is not directly coupled to the leads,
while finite correlations in the directly-coupled dot can further enhance this effect.
Moreover, we complement accurate numerical results
with a simple qualitative explanation based on 
analytical expressions for the zero-temperature conductance.
The proposed device provides a prospective example of an electrically-controlled,
fully spin-polarized current source, which operates without an external magnetic field.
\end{abstract}

\pacs{72.25.Mk, 73.63.Kv, 85.75.-d, 73.23.Hk}

\maketitle

\section{Introduction}
\label{sec:intro}

Efficient generation and control of spin currents at the nanoscale is 
one of the main goals of spin nanoelectronics \cite{book:awschalom02,
book:maekawa02,zuticRMP04,SeneorJPCM07,barnasJPCM08} .
This is because highly spin-polarized currents can be used to
address and detect the spin state of a magnetic nanostructure,
such as, e.g., a magnetic quantum dot or a single molecular magnet
\cite{timmPRB06,misiornyPRB07,lothNatPhys10} ,
which is of vital importance for applications in information storage
technologies. One of the easiest ways to generate high spin polarization
$\pol$ of the current is to apply an external magnetic field to the system.
If one considers then a singly occupied quantum dot,
the current becomes fully spin-polarized provided the transport voltage 
is smaller than the corresponding Zeeman splitting of the dot's level.
However, this method has two drawbacks: First,
the magnetic field needs to be strong enough to ensure that
$\pol \approx 1$ in a sufficiently large range of bias voltage,
which, however, can lead to undesired effects in the nanosystem, 
on which the spin-polarized current is to act. Second, changing 
the sign of $\pol$ requires a change in the direction of the magnetic field,
which in typical experiments cannot be realized at a rate comparable
to operations one would like to perform in a competitive spintronic device.

It has been recently shown that these disadvantages can be overcome
by using a quantum dot or a molecule strongly coupled to ferromagnetic 
leads \cite{csonka12}. The presence of ferromagnets results then in the occurrence 
of an exchange field, which leads to the splitting of the dot level similarly to 
an external magnetic 
field \cite{MartinekPRL03,Pasupathy04,HauptmannNatPhys08,GaassPRL11,
WeymannPRB11}. Now one obtains a splitting, whose sign and magnitude
can be controlled by a gate voltage, without any need to apply an external 
magnetic field. This splitting can lead to an enhancement of
the spin polarization. However, 
to reach full spin polarization, the system needs
to be highly left-right asymmetric \cite{csonka12,wojcikJPCM13}.

\begin{figure}[t]
\includegraphics[width=0.68\columnwidth,angle=270]{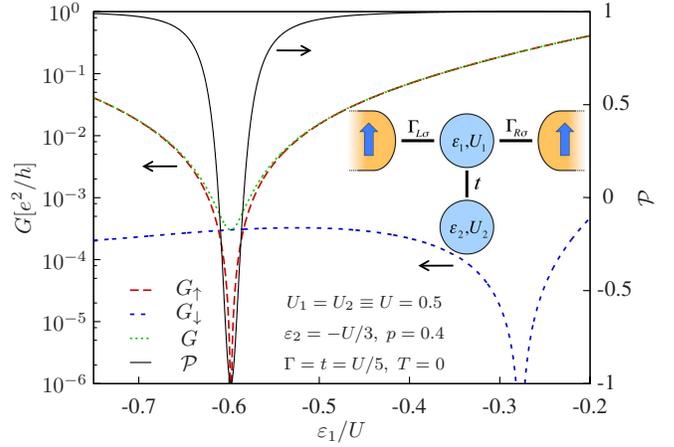}
\caption{(Color online)
     The spin-resolved linear conductance, $G_\sigma$, the total conductance,
     $G$, and the spin polarization, $\pol$, 
     obtained by the numerical renormalization group method,
     as a function of the first dot level $\e_1$ for typical DQD parameters indicated in the figure.
     The spin-dependent Fano effect leads to perfect spin polarization,
     the sign of which can be controlled by tuning the dot level position.
     See Sec.~\ref{sec:model} for details of the model and method.}
\label{fig:intro}
\end{figure}

In this paper, we propose a device with which one can induce
perfect spin polarization without an external magnetic field that can be
controlled by purely electrical means. The device does not need to be asymmetric either.
It consists of a double quantum dot (DQD) in a T-shaped geometry
coupled to external ferromagnetic leads. In this geometry, only one of the dots
is coupled directly to the leads, while the second dot is coupled indirectly,
through the first dot; see the inset of \fig{fig:intro}.
In T-shaped DQDs, the interference of different conduction paths
can lead to Fano antiresonance in the linear conductance
\cite{fanoPR61,sasakiPRL09,zitkoPRB10}.
In addition, the exchange field induced by the coupling to ferromagnets gives rise to the spin
splitting of the dots' levels. We will show that this leads 
to different conditions for destructive Fano interference in each spin channel.
As a result, there is a range of DQD's level positions where the difference
between the conductance in each spin channel is as large as a few orders of
magnitude and the spin polarization becomes essentially perfect.
This is illustrated in \fig{fig:intro}, which shows the linear conductance 
and the spin polarization as a function of the first dot level position for
typical DQD parameters indicated in the figure,
calculated by using the numerical renormalization group (NRG) method \cite{Wilson}.
The mechanism leading to 
$|\pol|\to 1$ is clearly visible: the spin-resolved conductance $G_\sigma$
displays Fano antiresonance at different $\e_1$. This gives rise to full spin
polarization, which changes sign just at the level position where $G$ is minimum.
Importantly, the whole operation is performed without 
any external magnetic field and can be controlled by only electrical means.

Recently, the transport properties of T-shaped DQDs coupled to nonmagnetic leads
have been analyzed by Dias da Silva {\it et al.} \cite{daSilvaPRB13} .
They focused on the role of external magnetic field and demonstrated
that such a system may work as a spin valve, 
producing spin polarization of the current $\P \approx \pm 1$ 
in an appropriately adjusted field. This effect also stems from
the spin-dependent Fano effect, in which the positions 
of Fano dips in respective spin channels are shifted with respect to each other.
Similar spin filtering effects have also been studied in 
transport through a quantum dot side-coupled to a quantum wire
\cite{torio_PRB02,torio_EPJB04,aligia_PRB04} .
In our device with ferromagnetic contacts,
we show that the same is possible without applying any magnetic field.
The spin polarization is then controlled by tuning the DQD's levels,
which is, no doubt, preferable from an application point of view. 
We note that the transport properties of T-shaped DQDs with ferromagnetic contacts
have already been addressed in a few papers \cite{yang07,yang08,hou09,rui12}.
These considerations were, however, restricted to a rather weak-coupling
regime, and the effects of the exchange field were not properly taken into
account. Our analysis is performed with the aid of NRG, which allows us to
study the effects related to a ferromagnetic-contact-induced exchange 
field in a very accurate way.

This paper has the following structure:
Having introduced the model and method in Sec.~\ref{sec:model},
in Sec.~\ref{sec:P} we discuss
the behavior of the spectral function determining the linear conductance, and we 
explain the physical reasons for the occurrence of enhanced spin polarization in the
system. We also provide approximate analytical
formulas for the exchange field, which agree well with the NRG results.
Finally, we present the results of NRG calculations for the linear conductance and the spin 
polarization in Sec.~\ref{sec:NRG}, and we conclude the paper in Sec.~\ref{sec:end}.

\section{Model and method}
\label{sec:model}

We consider a double quantum dot forming a T-shaped configuration coupled to ferromagnetic
leads whose magnetizations are oriented in parallel; see the inset of
Fig.~\ref{fig:intro}. The first dot is coupled directly to the left (right) lead
with coupling strength $\Gamma_{L\sigma}$ ($\Gamma_{R\sigma}$), while
the second dot is coupled to the first one through the hopping
parameter $t$. The Hamiltonian of the system has the form
\beq
H = H_{\rm F} + H_{\rm T} + H_{\rm DQD},
\label{Hform}
\eeq
where 
\beq
H_{\rm F} = \sum_{r=L,R} \sum_{\bk\s} \e_{r\bk\s} c^\dag_{r\bk\s}c_{r\bk\s}
\label{HF}
\eeq
is the Hamiltonian of ferromagnetic leads treated in a noninteracting particle
approximation, the tunneling Hamiltonian is given by
\beq
H_{\rm T} = \sum_{r=L,R} \sum_{\bk\s} V_{r\bk\s} \left(\dk_{1\s} c_{r\bk\s} +c_{r\bk\s}^\dag d_{1\s}  \right) \, ,
\label{HT}
\eeq
and the DQD Hamiltonian reads, 
\beqa
H_{\rm DQD} &=& \sum_{j\s} \e_{j\s} d^\dag_{j\s}d_{j\s}  + \sum_j U_j d^\dag_{j\up} d_{j\up}d^\dag_{j\down} d_{j\down} \nn\\
	&&+\, t \sum_\s (d^\dag_{1\s}d_{2\s} + d^\dag_{2\s}d_{1\s}) \, .
	\label{HQD}
\eeqa
Here, $d_{j\s}$ annihilates an electron with spin $\s$ on dot $j$, 
$c_{r \bk \s}$ annihilates an electron with spin $\s$ and momentum $\bk$ 
in lead $r$, $\e_{j\s}$ and $\e_{r\bk \s}$ denote the energies of respective
electrons, $U_j$ is the Coulomb interaction on dot $j$ and $V_{r\bk\s}$ 
denotes the corresponding tunnel matrix element. The spin-dependent coupling 
to the contact $r$ is given by 
$\G_{r\s} = \sum_{\bk} \pi \rho_{r\s} |V_{r\bk\s}|^2$, where $\rho_{r\s}$ is 
the spin-dependent, normalized density of states of lead $r$. Here, we model 
the coupling by $\G_{r\s} = (1 +\s p) \G_r$, where $p$ is the spin polarization
of the ferromagnets and $\G_r = (\G_{r\uparrow} + \G_{r\downarrow})/2$.
In the following, we assume $\G_L = \G_R \equiv \G/2$. We also assume that the
Coulomb correlations between the two dots are very weak and can be neglected.
We use the band half-width as the energy unit, $D\equiv 1$.

The linear-response conductance in spin channel $\s$ can be found 
from \cite{Meir-Wingreen}
\beq
G_\s = \frac{e^2}{h} \G_\s
  \int \de \omega \frac{\dd f(\omega)}{\dd\omega} \,\Im\! \GF{ d_{1\s} | \dk_{1\s} }^{\! \rm ret}_\omega \, ,
  \label{M-W}
\eeq
where $\G_\s = \G_{L\s} + \G_{R\s}$, $f(\omega)$ is the Fermi-Dirac distribution
function and $\GF{ d_{1\s} | \dk_{1\s} }^{\!\rm ret}_\omega$ denotes the Fourier
transform of the retarded Green's function of the first quantum dot.

To obtain reliable results of high accuracy for our strongly interacting system,
we employ the numerical renormalization group method \cite{Wilson}.
By using the complete eigenbasis of the NRG
Hamiltonian, we construct the thermal density matrix of the 
system \cite{anders05,weichselbaumPRL07}, which allows us to calculate various
correlation functions at arbitrary temperatures.
Here, to perform calculations, we use the Budapest Flexible DM-NRG code
\cite{fnrg,tothPRB08}.

The main quantity in which we are interested, apart from linear conductance, is 
the spin polarization, which is defined as
\beq
\P \equiv \frac{G_\up - G_\down}{G_\up + G_\down}.
\label{P-def}
\eeq
At zero temperature, formula (\ref{M-W}) simplifies considerably to,
$G_\s = (e^2 / h) \pi \G_\s A_{1\s} (0)$,
where 
$A_{1\s}(\w) = - \Im\! \GF{ d_{1\s} | \dk_{1\s} }^{\! \rm ret}_\omega / \pi$
denotes the spin-resolved spectral function of the first dot. Then, the spin
polarization can be expressed in terms of the normalized spectral function,
$\A_{1\s}(\omega) = \pi\G_\s A_{1\s}(\omega)$, taken at $\omega=0$, as 
$\pol = [\A_{1\up}(0) - \A_{1\down}(0)] / \A_{1}(0)$, with 
$\A_1(\w) = \sum_\s \A_{1\s}(\w)$.

\section{Origin of enhanced spin polarization}
\label{sec:P}

Since the linear conductance and the spin polarization are expressed in terms
of the first dot spectral function, we will focus on its behavior.
To understand the origin of large spin polarization in the considered system,
we first consider the case of noninteracting T-shaped DQD and then 
study the effect of Coulomb correlations.

\subsection{Noninteracting case}

For $U_1 = U_2 = 0$, with the aid of the equation of motion,
the spectral function of the first dot can be expressed as
\beq
A_{1\s}(\omega) = \frac{1}{\pi} \frac{\G_\s}{\left[ \omega - \e_{1\s} 
	- t^2/(\omega-\e_{2\s})  \right]^2 + \G_\s^2}.
\eeq
Then, the spin-resolved linear conductance at zero temperature is given by
\beq 
  G_\s = \frac{e^2}{h}  \frac{\G_\s^2} {(\e_{1\s} - t^2 / \e_{2\s})^2 + \G_\s^2}.
  \label{eq:Gs}
\eeq
Let us now consider some limiting cases. For nonmagnetic leads, $p=0$, and in 
the absence of magnetic field, $\e_{j\s} = \e_j$, the linear conductance at 
$T=0$ is given by
\beq
  G = \frac{2 e^2}{h} \frac{\G^2} {(\e_1 - t^2/\e_2)^2 + \G^2},
\eeq
which for $\e_1=0$ yields
\beq
  G = \frac{2 e^2}{h} \frac{E^2} {1 + E^2},
\eeq
with $E = \e_2 / \G_2$ and $\G_2 = t^2 / \G$.
This is the well-known Fano formula describing symmetric antiresonance
as a function of energy $E$ \cite{fanoPR61,MiroshnichenkoRMP10}.
For $\e_1 = 0$, the half-width of the minimum in $G$ is given by $t^2/\G$.
When $\e_1 \neq 0$, the antiresonance is still located 
at $\e_2 = 0$, however, it becomes asymmetric \cite{zitkoPRB10}.

In the presence of an external magnetic field $B$,
the position of the Fano antiresonance depends on spin, see \eq{eq:Gs},
since it occurs at $\e_{2\s} = \e_2 + \s B/2 = 0$, where 
$g\mu_B \equiv 1$. Consequently, while for one spin direction
the conductance is finite, for the other one it can be fully suppressed,
leading to $|\pol| = 1$.
Assuming $p=0$ and $\e_{1\s}  = \e_{2\s} = \e + \s B/2$, 
the spin polarization is then given by
\beq
  \P \!=\! \frac{\e B [t^4 - (\e^2-B^2/4)^2]}{(\e^2\!+\!B^2/4)t^4 \!+\! (\e^2\!-\!B^2/4)^2 (\e^2 \!+\! B^2/4 \!+\! \G^2 \!-\! 2t^2)}.    
\label{P-U0-B}
\eeq
For $\e=B/2$, one has, $\P=1$, while for $\e = -B/2$, $\P=-1$.
Thus, for finite $B$, the spin polarization can be enhanced to its
maximum value, and its sign can be changed, depending on the DQD's levels.
This effect is completely destroyed in $B=0$, unless $p\neq 0$.
In the case of ferromagnetic leads and in the absence of magnetic field
(henceforth we assume $\e_{j\s} \equiv \e_j$),
for the spin polarization of the linear conductance, one finds 
\beq
   \P = \frac{2p}{1+p^2} \frac{(\e_1 - t^2/\e_2)^2}{(\e_1 - t^2/\e_2)^2 + (1-p^2)^2\G^2 / (1+p^2)} .
\label{P-U0-p}
\eeq
From this formula, it follows that $\pol = 0$ for $\e_1\e_2 = t^2$
and $\pol = 2p^2 / (1+p^2)$ for $\e_2 = 0$, irrespective of $\e_1$.
Thus, the spin polarization is finite, $0 \leq \pol \leq 2p/(1+p^2)$,
but it does not change sign and is always smaller than unity for $p<1$.

\begin{figure}[t]
\centering
\includegraphics[width=1\columnwidth]{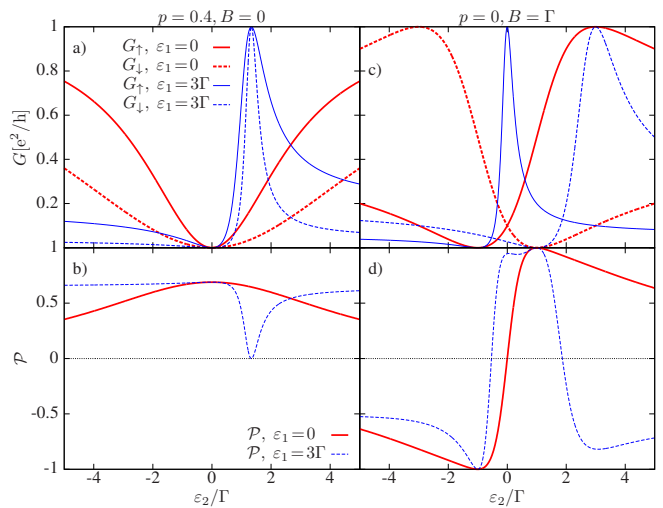}
\caption{(Color online)
             The spin-resolved linear conductance (first row) and the spin polarization (second row)
             as a function of $\e_2$ for two values of $\e_1$, as indicated, and for $t=2\G$
             in the case of noninteracting DQDs.
             The left column corresponds to $p=0.4$ and $B=0$, while
             the right column to $p=0$ and $B=\G$.}
\label{fig:noint}
\end{figure}

The spin-resolved conductance and the spin polarization
for noninteracting dots are plotted in \fig{fig:noint}.
In the absence of magnetic field, for $\e_1=0$, the linear conductance
in each spin channel displays a symmetric antiresonance
as a function of $\e_2$ located at $\e_2 = 0$,
which becomes asymmetric once $\e_1 \neq 0$; see \fig{fig:noint}(a).
The spin polarization is also asymmetric for $\e_1\neq 0$.
Moreover, $\pol$ is positive in the whole range of $\e_2$
and becomes fully suppressed for $\e_2/\G = t^2 / (\e_1\G) = 4/3$; see \fig{fig:noint}(b).
At this point, the linear conductance reaches its maximum value,
irrespective of spin channel $\s$.
In the case of a finite magnetic field and nonmagnetic leads, the Fano antiresonance
is asymmetric in each spin channel even for $\e_1 = 0$ [\fig{fig:noint}(c)],
and the minimum in $G_\s$ occurs at different $\e_2$.
This leads to full spin polarization $\pol$, which can change sign
in a certain range of $\e_2$; see \fig{fig:noint}(d).
Figure \ref{fig:noint} clearly demonstrates the difference between the two cases
discussed above. In the case of noninteracting dots, spin-dependent tunneling
due to $\G_\up\neq\G_\down$ (in the absence of $B$)
does not lead to a spectacular dependence of $\pol$ on the DQD's levels.

\subsection{Interacting case}
\label{sec:P-int}

The spin polarization of the T-shaped DQD with ferromagnetic contacts
for $B=0$ can be enhanced considerably when one includes the interactions
in the dots. For finite $U_1$ and $U_2$, the Green's function of the first 
dot is given by
\beq
  \GF{ d_{1\s} | \dk_{1\s} }^{\!-1}_{\w} \!\! = \w - \e_{1}  - \Sigma_{1\s}\!(\w)- \frac{t^2}{\w\!-\!\e_{2}\!-\!\Sigma_{2\s}\!(\w)}  + i \G_\s ,
\eeq
where the self-energy $\Sigma_{j\s}$ is defined as
\beq
  \Sigma_{j\s}(\w) = U_j \frac{ \GF{d_{j\s}n_{j\sbar} | \dk_{1\s}}_\w }{ \GF{d_{j\s}|\dk_{1\s}}_\w } \;\;\;\;\; (\sbar \equiv -\s).
\eeq
One can now use the equation-of-motion technique to find the higher-order 
Green's functions and solve the problem by using an appropriate decoupling scheme.
This is, however, not the goal of our paper, since we calculate the Green's
functions by NRG, which enables us to obtain very accurate results.
Nevertheless, to get some intuitive understanding of what happens
in correlated T-shaped DQDs, let us consider the zero-temperature conductance
(note that for $\w=0$, the self-energy is real),
\beq 
  G_\s \!= \!\frac{e^2}{h} \frac{ \G_\s^2 }
  {\{ \e_1 + \Sigma_{1\s}(0) - t^2 /[\e_2 + \Sigma_{2\s}(0)] \}^2 + \G_\s^2}.
  \label{eq15}
\eeq
Then, we employ the simplest mean-field approximation to the self-energies,
$\Sigma_{j\s} \approx U_j \langle n_{j\sbar}\rangle$, which
allows us to extract a few interesting conclusions from \eq{eq15}.
The most important one is that when $\e_2 + 
U_2\la n_{2\bar{\s}} \ra = 0$, the conductance in 
spin channel $\s$ becomes suppressed due to the Fano destructive
interference. If $\la n_{2\up}\ra \neq \la n_{2\down}\ra$, the 
conditions for conductance suppression are different in 
each spin channel. The spin imbalance in dot level occupation 
can be induced by the presence of an exchange field, as described 
in the following subsection.

The difference in the positions of Fano antiresonances for different
spin channels is illustrated in Fig.~\ref{fig:intro}.
Indeed, $G_\up$ has a minimum for different $\e_1$
compared to $G_\down$, and the resulting $\P$ reaches $\pm 1$.
Moreover, it can be observed that $\P$ changes sign at the level position
for which the total conductance is minimum.

The second significant conclusion, which can be drawn from Eq.~(\ref{eq15}),
is that it is sufficient to have different occupations
for given spin only in the second dot. This implies
that the first dot does not need to be interacting.
Finally, the enhanced spin polarization occurs when the 
second dot is in the local moment regime, $-U_2 < \e_2 < 0$,
while no such restriction is imposed on the first dot.

\subsection{Exchange field}

The coupling to external leads gives rise to renormalization of the DQD's levels.
Since in the case of ferromagnetic leads the coupling $\G_\s$
depends on spin direction, the level renormalization is also spin-dependent.
This results in spin-splitting of the levels,
$\exch^{(j)} = \delta \e_{j\up} - \delta {\e}_{j\down}$,
where $\exch^{(j)}$ is the exchange field on dot $j$ and 
$\delta \e_{j\s}$ denotes the respective spin-dependent level renormalization.

Contrary to the Zeeman splitting caused by an external magnetic field, the sign and
magnitude of the splitting induced by ferromagnetic leads can be tuned by changing 
the position of the quantum dot levels. \cite{MartinekPRL03,MartinekPRB05} 
To understand the effect of an exchange field on transport through T-shaped DQDs,
we will consider some limiting situations.
In the case of $t=0$, the exchange field on the first dot
can be found within the perturbation theory, which in the second order gives
\cite{MartinekPRL03,MartinekPRB05} 
\beq
\exch^{(1)} = \frac{2p\G}{\pi} \log \left|\frac{\e_1}{\e_1+U_1} \right|. 
\label{eq:exch}
\eeq
Note that $\exch^{(1)}$ clearly results from correlations and vanishes for
$U_1=0$. Moreover, it also vanishes at the particle-hole symmetry point,
$\delta_1 = 0$, with $\delta_j = \e_j + U_j/2$, denoting the detuning of
dot $j$ from the symmetry point.

\begin{figure}[t]
\centering
\includegraphics[width=1\columnwidth]{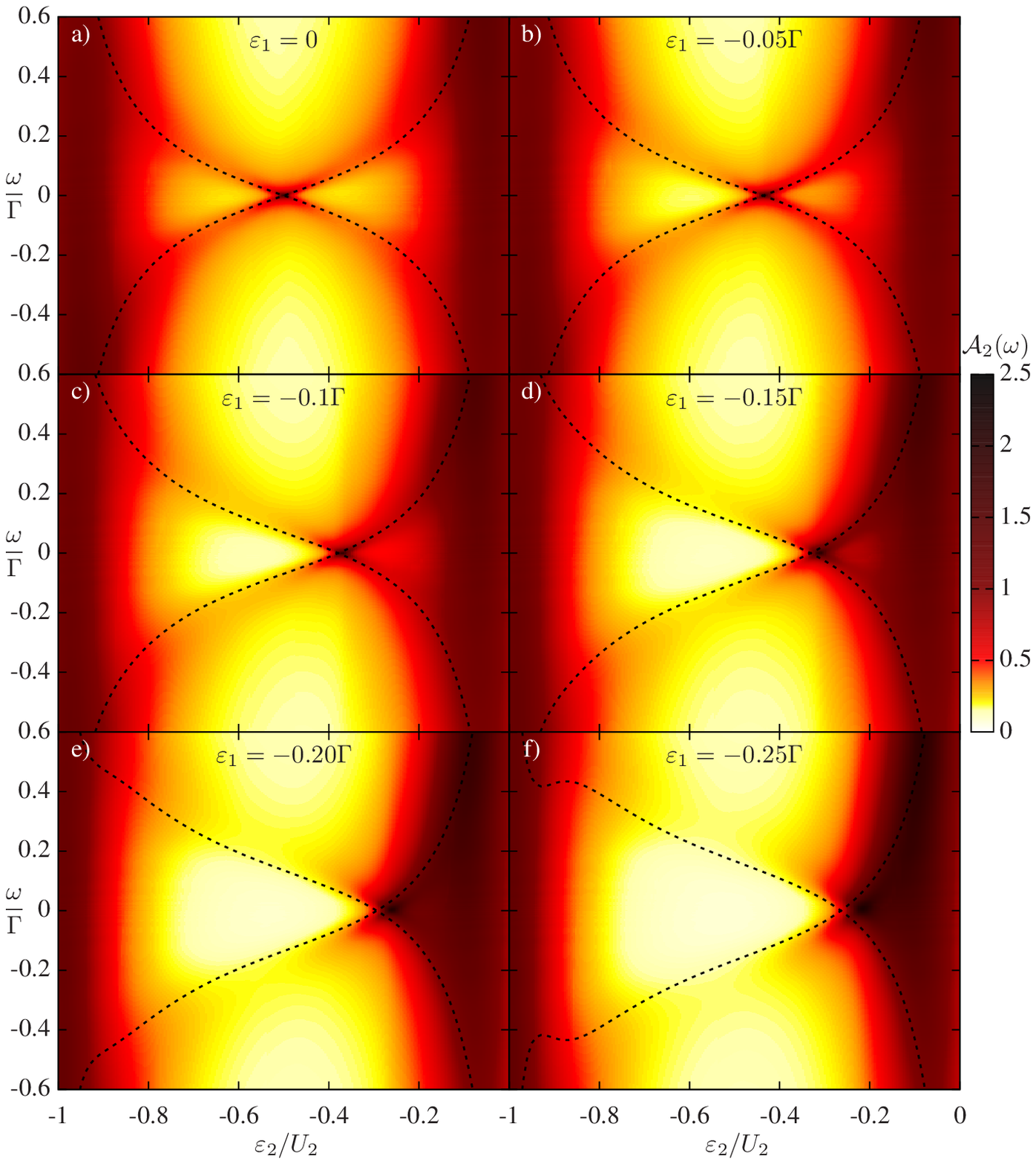}
\caption{(Color online)
             The normalized spectral function of the second dot
             $\A_2(\w)$ plotted as a function of energy $\omega$
             and position of the second dot level $\e_2$
             for (a) $\e_1 = 0$, (b) $\e_1 = -0.05\G$,
             (c) $\e_1 = -0.1\G$, (d) $\e_1 = -0.15\G$,
             (e) $\e_1 = -0.2\G$, and (f) $\e_1 = -0.25\G$.
             The dashed lines present the results obtained from
             analytical formula (\ref{Eq:exch2a}).
             The parameters are: $U_1 = 0$, $U_2=U=0.5$,
             $\Gamma=U/5$, $t=\Gamma/2$ and $p=0.4$.}
\label{fig:exch}
\end{figure}

Now, let us see what happens in the second dot. Since,
as follows from previous discussion, to obtain large spin
polarization it is sufficient to have interactions only in the second dot, 
we now assume $U_1=0$. The hybridization of the second dot depends 
on the local density of states of the first dot, 
$\G_{2\s}(\w) = \pi A_{1\s}^0(\w) t^2$, where $A_{1\s}^0(\w)$ denotes the spectral
function of the first dot in the case of $t=0$,
\beq
A_{1\s}^0(\w) = \frac{1}{\pi} \frac{\Gamma_\s}{(\w-\e_1)^2 + \Gamma_\s^2}.
\eeq
In this way, the model becomes equivalent to the Anderson model with a Lorentzian 
density of states.
Since the leads are ferromagnetic, $A_{1\s}^0(\w)$ depends on spin through $\G_\s$,
and so does $\G_{2\s}(\w)$, which for low energies ($\w= 0$) and $\e_1=0$ becomes
equal to $t^2/\G_\s$.
Note, that the dependence of couplings on spin is opposite in each dot:
while $\G_\up > \G_\down$, for the
second dot the spin-down level is more strongly coupled than the spin-up one,
$\G_{2\up}(0) < \G_{2\down}(0)$. In the second order of perturbation theory,
renormalization of the second dot's level is given by
\beq
\delta \e_{2\s} = \frac{1}{\pi} \! \int \!\! \de\w \! \left[ \frac{\G_{2\s}(\w)   f^-(\w)}{\e_{2} - \w} 
				+ \frac{\G_{2\bar{\s}}(\w) f(\w)}{\w - \e_{2} - U_2} \right] ,
\label{eq:exInt}
\eeq
where $f^{-}(\w) = 1-f(\w)$. When assuming the limit of zero temperature,
taking $\e_1=0$, and approximating the hybridization by $\G_{2\s}(\w) = t^2/\G_\s$,
for the exchange field $\exch^{(2)}$ one finds
\beq
\exch^{(2)} = -\frac{t^2}{\pi\G}\frac{2p}{1-p^2} 
	\log \left|\frac{\e_2}{\e_2 + U_2} \right|.
\label{eq:exch2}
\eeq
Although this formula is very simplified, it still
allows us to correctly extract the intuitive behavior of the system. First of all,
one can see that the presence of ferromagnets is also revealed in the second dot. 
It leads to the exchange field, which has a similar dependence on the level position,
in the way that it vanishes for $\delta_2 = 0$, but it has a different magnitude and sign
(for given detuning) compared to $\exch^{(1)}$; cf. \eq{eq:exch}. Thus, if one
would like to mimic the effect of external magnetic field by the exchange field,
the detuning in each dot should have an opposite sign. However, it is worth stressing
that the exchange field offers much more flexibility, since it allows the
spin-splitting to be tuned in each dot separately by gate voltages.
For completeness, we also present the zero-temperature formula for the exchange field $\exch^{(2)}$
in the case of $\e_1\neq 0$ and for energy-depednent
hybridization $\Gamma_{2\s}(\w)$. It is given by
\beqa
\exch^{(2)} \es \sum_\s \s \frac{t^2}{2} \big[ L_{\G_\s}(U_2\smin \Delta) - L_{\G_\s}(\Delta) \big] \nn\\
	&-& \sum_\s \s \frac{t^2}{\pi} \arctan\left( \frac{\e_1}{\G_\s} \right) 
		\big[ L_{\G_\s}(U_2\smin \Delta) + L_{\G_\s}(\Delta) \big] \nn\\
	&-& \sum_\s \s \frac{t^2}{2\pi}  L_{U_2\smin \Delta}(\G_\s)\log \frac{(\e_2+U_2)^2}{\e_1^2 + \G_\s^2} \nn\\
	&+& \sum_\s \s \frac{t^2}{2\pi}  L_{\Delta}(\Gamma_\s) \log \frac{\e_2^2}{\e_1^2 + \G_\s^2}  , 
\label{Eq:exch2a}
\eeqa
where $L_y(x) = x/(x^2 + y^2)$ and $\Delta = \e_1 - \e_2$.

We note that in the case of a noninteracting first dot,
the model corresponds to the single-impurity Anderson model
with nonconstant density of states. At low temperatures,
one should then expect a single-stage Kondo effect to occur
\cite{kondo64,Kastner:1998km,Cronenwett:1998hy,daSilvaPRB13,
daSilvaPRL06,VaugierPRB07,daSilvaPRB08}.
However, due to the presence of the exchange field,
the Kondo resonance becomes suppressed,
which happens once $|\exch^{(2)}|\gtrsim T_K$,
where $T_K$ is the Kondo temperature.
Thus, for T-spahed DQDs with ferromagnetic contacts,
the Kondo effect is generally suppressed. 
In \fig{fig:exch} we show the NRG results on the normalized
spectral function of the second quantum dot,
$\A_2(\w) = \sum_\s \pi t^2 A_{2\s}(\w) / \G_\s $, where 
$A_{2\s}(\w)$ denotes the spectral function of the second dot.
For $\e_1 = 0$, at the particle-hole symmetry point, $\delta_2=0$,
the effect of the exchange field is negligible
and the spectral function exhibits Kondo resonance
\cite{kondo64,Kastner:1998km,Cronenwett:1998hy}.
The Kondo temperature, defined as the half-width at half-maximum
of the Kondo peak in the spectral function, for parameters assumed in 
\fig{fig:exch} and for $\e_1=0$ and $\delta_2=0$, is equal to, $T_K\approx 0.005\Gamma$.
When $\delta_2 \neq 0$ and $|\exch^{(2)}|\gtrsim T_K$,
the exchange field leads to the spin splitting of the Kondo resonance; see \fig{fig:exch}(a).
We note that such a splitting of the Kondo effect due to the presence of ferromagnets
has already been observed experimentally in single quantum dots
\cite{Pasupathy04,HauptmannNatPhys08,GaassPRL11}.
When $\e_1\neq 0$, the splitting of the Kondo resonance becomes asymmetric
around $\delta_2 = 0$, and the point where the exchange field is suppressed moves
towards the resonance at $\e_2 = 0$, until it actually merges with the resonant peak. 
One observes then a spin splitting whose magnitude can be tuned,
but the sign does not change, see \fig{fig:exch}.
In the case of $|\exch^{(2)}|\gtrsim T_K$ the Kondo peak is split and the 
spectral function shows only side resonances, which occur at $\omega = \pm |\exch^{(2)}|$.
\cite{GaassPRL11}
The dashed lines in \fig{fig:exch} show the positions
of these resonances based on \eq{Eq:exch2a}. 
As can be seen, they match nicely with the numerical data 
for all values of $\e_1$ presented in the figure.

As follows from the above discussion, the effective exchange field
induced by the presence of ferromagnets
can be conveniently tuned by sweeping the gate voltages
and adjusting the positions of the DQD's levels.
This is of importance from an experimental point of view.
We also note that in general the splitting of the Kondo peak can also occur
in the case of relatively large hopping between the dots.
\cite{daSilvaPRL06,VaugierPRB07,daSilvaPRB08}
However, for parameters assumed in \fig{fig:exch}, such splitting is absent.\cite{VaugierPRB07}
The observed splitting is exclusively due to the presence of the exchange field.

\section{Numerical results}
\label{sec:NRG}

In the following, we present and discuss the numerical results
on the spin-resolved linear conductance $G_\s$ and the spin polarization $\pol$.
Previous discussion showed that for the full spin polarization to occur,
it is necessary to have interactions in the second dot,
while the first dot can be noninteracting. Therefore, we first study
the case of $U_1 = 0$ and finite $U_2$, and then we also include
the interactions in the first dot and analyze how they influence
the linear conductance and the spin polarization of the system.
Finally, we discuss the effect of finite temperature on transport properties.

We also note that to observe an enhanced spin polarization and tune its sign,
one can fix the level of one of the dots and tune the other one.
Since it is crucial to have an exchange field in the second dot,
we thus fix the level of the second dot, such that $\delta_2 \neq 0$, and tune
the position of the first dot. (This is what is presented in \fig{fig:intro}
for a general interacting case.) Nevertheless, we also present
the density plot of the spin polarization as a function of both 
$\e_1$ and $\e_2$.

\subsection{The case of noninteracting first dot}
\label{sec:NRG-nonInt}

\begin{figure}[t]
\centering
\includegraphics[width=1\columnwidth]{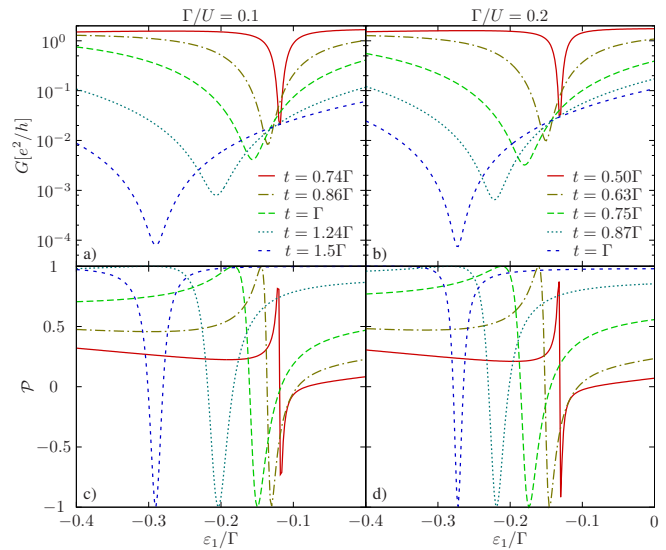}
\caption{(Color online)
             The linear conductance (first row) and the spin polarization (second row)
             as a function of $\e_1$ for $\G/U=0.1$ (left column)
             and $\G/U=0.2$ (right column) calculated for different values
             of the hopping $t$ between the dots, as indicated.
             The parameters are $U_1=0$, $U_2=U=0.5$, $\e_2 = -U/3$, $p=0.4$
             and $T=0$.}
\label{fig:Gt}
\end{figure}

The total linear conductance and spin polarization in the case of
$U_1=0$ and $U_2=U=0.5$ are shown in \fig{fig:Gt}
for two values of the coupling $\G$ and for different hoppings $t$ between the dots.
The position of the second dot level is $\e_2 = -U/3$,
to assure that the exchange field effects are present in the system.
Since the strength of the exchange field is proportional
to $\G_2=t^2/\G$, cf. \eq{eq:exch2}, by increasing $t$, one also
increases the magnitude of the exchange-field-induced spin splitting
of the second dot's level. As a consequence, the conditions for destructive 
interference change in each spin channel with tuning $t$
and the dependence on $t$ is different for each coupling $\G$;
see \fig{fig:Gt}.

First of all, one can see that by increasing $t$, the total conductance
decreases. For large $t$, [see, e.g., the case of $t=1.5\G$ in \fig{fig:Gt}(a)
or $t=\G$ in \fig{fig:Gt}(b)], the conductance is three or four orders
of magnitude smaller than the conductance quantum.
Although these values are rather small, they are still measurable experimentally.
In fact, similar values of $G$ occur in quantum dots in the cotunneling regime
\cite{koganPRL04}.
For $\e_1$, where $G$ takes its minimum value, the spin polarization
changes sign and becomes negative. This sign change becomes
enhanced upon increasing the exchange field (increasing $t$),
and for large $t$, the spin polarization becomes perfect
and changes sign from $+1$ into $-1$. Thus, for given $t$ and nonzero
detuning $\delta_2\neq 0$, the spin polarization
can be tuned by only electrical means, namely by shifting
the position of the first dot level with a gate voltage.
The role of the exchange field is crucial here, which can be deduced 
from the fact that the effect disappears for $\delta_2 = 0$, when 
$\exch^{(2)} = 0$; cf. \eq{eq:exch2}.

\begin{figure}[t]
\centering
\includegraphics[width=\columnwidth]{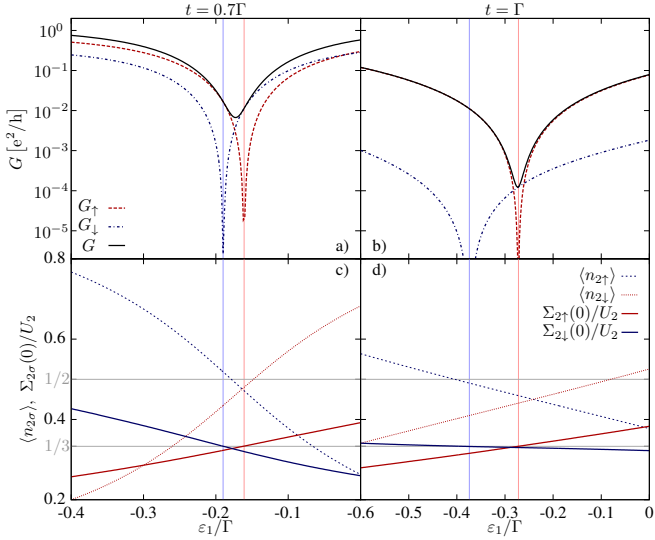}
\caption{(color online)
             The spin-resolved and total conductances (a)-(b),
             and the spin-dependent occupations
             together with the self-energies for $\omega=0$ (c)-(d)
             of the side-coupled quantum dot as a function of $\e_1$
             calculated by NRG for $\G/U=0.2$, $t=0.7\G$ (left column) and $t=\G$ (right 
             column). The vertical dotted lines mark the positions where the minima
             in spin-dependent conductance occur.
             The horizontal lines correspond to $-\e_2/U_2 = 1/3$ and $1/2$.
             The minimum of $G_\s$ occurs at the crossing of $\Sigma_{2\s}(0)/U_2$ with $1/3$,
             cf. \eq{eq15}. The Friedel sum rule predicts the minimum to occur
             when $\la n_{2\s}\ra = 1/2$.
             The other parameters are the same as in \fig{fig:Gt}.}
\label{fig:n}
\end{figure}

The minimum in $G_\sigma$ occurs for such $\e_1$ that
$\Sigma_{2\s}(\omega\!=\!0) = -\e_2$; cf. \eq{eq15}.
This is explicitly illustrated in \fig{fig:n}, which shows
the spin-dependent conductance, $G_\s$, and self-energy for $\w=0$,
$\Sigma_{2\s}(\omega\!=\!0)$, and a function of
$\e_1$ for two different hoppings: $t=0.7\G$ (left column) and $t=\G$ (right column).
Since in calculations we assumed $\e_2/U_2=-1/3$,
the minimum in $G_\s$ occurs precisely at the point
where $\Sigma_{2\s}(\omega\!=\!0) = 1/3$; see \fig{fig:n}.

As mentioned in the preceding section, for $U_1=0$ the model is equivalent to the 
single-impurity Anderson model with a Lorentzian density of states. Then, the Friedel 
sum rule allows one to relate the conductance through the system to the spin-resolved
occupation of the second dot \cite{FSRforSIAM}. For $t \ll \Gamma$, it can be written 
as $G_\s = (e^2/h)\, \cos^2(\pi \la n_{2\s}\ra)$ \cite{aligia_PRB04}. Thus, the
conductance in spin channel $\s$ should be suppressed when $\la n_{2\s} \ra = 1/2$.
However, for stronger hoppings, $t \sim \Gamma$,
the condition $\la n_{2\s} \ra = 1/2$ is not necessarily fulfilled
and the application of the Friedel sum rule becomes more complicated.
The spin-resolved occupations of the second dot as a function of $\e_1$
are shown in Figs. \ref{fig:n}(c) and \ref{fig:n}(d).
The critical occupation for which the conductance becomes minimum is 
still of the order of $1/2$, but its precise value is different.
On the other hand, for larger values of $t$, the phase shift, which determines the
position of the conductance minimum, is given by a rather complex
expression even in the particle-hole symmetry point \cite{daSilvaPRB13}.
In the case of significant particle-hole symmetry breaking,
as considered in this paper (note that this is a necessary condition
to have the exchange field present in the system),
it is very difficult to utilize the Friedel sum rule, 
nevertheless, the condition $\Sigma_{2\s}(\omega\!=\!0) = -\e_2$ is always correct
as long as $T=0$.
Finally, one can notice that the simplest mean-field
approximation used in Sec.~\ref{sec:P}, $\Sigma_{2\s} = U_2 \la n_{2\sbar} \ra$,
leading to the condition $\la n_{2\sbar} \ra = -\e_2/U_2$ for the minimum in $G_\s$
is also violated; see Figs. \ref{fig:n}(c) and \ref{fig:n}(d).
However, the qualitative analysis of the system behavior
based on this approximation is still sound.

\begin{figure}[t]
\centering
\includegraphics[width=0.8\columnwidth]{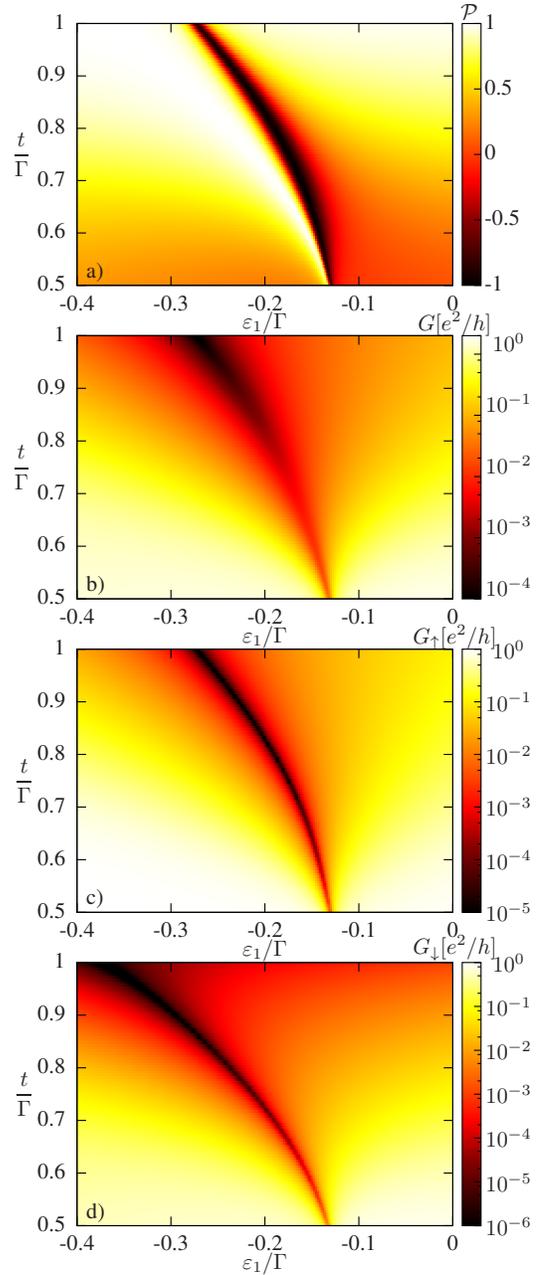}
\caption{(Color online)
             The spin polarization $\pol$ (a) and the
             logarithm of the linear conductance $G$ (b), $G_\up$ (c), $G_\down$ (d)
             as a function of $\e_1$ and $t$ calculated for
             parameters the same as in \fig{fig:Gt} with $\G/U=0.2$.}
\label{fig:Gt2D}
\end{figure}

The explicit dependence of the linear conductance and spin polarization
on both $\e_1$ and $t$ is shown in \fig{fig:Gt2D} for $\G/U=0.2$.
The conductance is plotted in logarithmic scale to indicate the position
of the conductance minimum due to the Fano effect.
Clearly, the minimum occurs at different level position in each 
spin channel, see Figs. \ref{fig:Gt2D}(c) and \ref{fig:Gt2D}(d).
Moreover, the spin-up conductance is generally much
larger than the spin-down conductance, except for the level position
where $G_\up$ is suppressed by the Fano effect.
Consequently, for this level position, the total conductance
has a minimum [\fig{fig:Gt2D}(b)], while the spin polarization
changes sign and becomes $\pol \approx -1$, otherwise $\pol \approx 1$;
see \fig{fig:Gt2D}(a).
Note also that the position of the minimum in $G_\s$ occurs at different
$\e_1$ for different $t$, which results directly from the dependence
of the exchange field on $t$.

\subsection{Fully interacting case}
 
\begin{figure}[t]
\centering
\includegraphics[width=0.9\columnwidth]{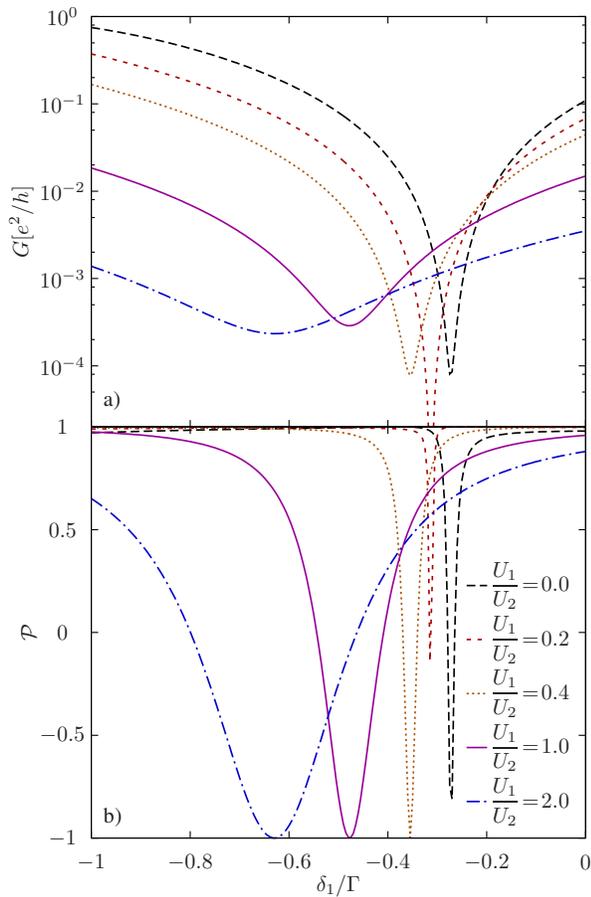}
\caption{(Color online)
			 The linear conductance (a) and the spin polarization (b)
             as a function of the first dot detuning $\delta_1$ calculated for 
             different Coulomb correlations in the first dot, as indicated.
             The other parameters are:
             $U_2=0.5$, $\G=t=U_2/5$, $\e_2=-U_2/3$, $p=0.4$, and $T=0$.}
\label{fig:GU}
\end{figure}

Let us now include the interactions in the first dot, $U_1\neq 0$.
The linear conductance and spin polarization as a function of the first dot
detuning $\delta_1$ for different correlations $U_1$ are shown in \fig{fig:GU}.
This figure is calculated for $U_2=0.5$, $\G = t = U_2/5$, and $\e_2=-U_2/3$.
For finite $U_1$ and $\delta_1\neq 0$, the exchange field also develops
in the first dot, cf. \eq{eq:exch}. We note that treating the exchange field
in each dot separately is mainly to increase the intuitive understanding
of the physics. However, we need to stress that for larger hoppings,
$t \gtrsim \G$, transport occurs through molecular many-body states of the DQD,
and formulas (\ref{eq:exch2}) and (\ref{Eq:exch2a}) based on perturbation theory in $t$
present only very crude estimations.

\begin{figure}[t]
\centering
\includegraphics[width=0.85\columnwidth]{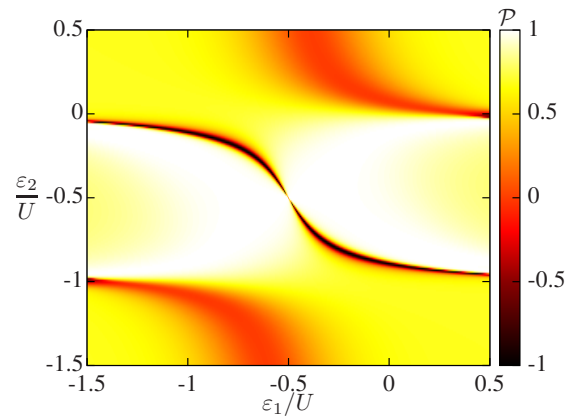}
\caption{(Color online) 
	 	  The spin polarization as a function of the DQD levels $\e_1$ 
          and $\e_2$ calculated for $U_1=U_2=U=0.5$. 
          The other parameters are the same as in \fig{fig:GU}.}
\label{fig:GU2D}
\end{figure}

By increasing $U_1$, the exchange field effects become generally enhanced.
It can be seen that the minimum in $G$ and $\pol$ as a function of $\delta_1$
changes position with $U_1$; see \fig{fig:GU}.
Moreover, the width of both the conductance minimum
and the spin-polarization sign change also increases with increasing $U_1$.
For example, when $U_1=U_2$, both $G$ and $\pol$ exhibit
an approximately symmetric minimum as a function of $\delta_1$.
Interestingly, for $U_1 = U_2/5$, the effect of the spin-polarization sign change
is weakened, while the conductance suppression is then very large.
For these parameters, the conditions for the Fano effect in each spin channel
become roughly equal, and the minimum in $G_\s$ occurs at comparable $\delta_1$
in each spin channel.
We also note that for positive detuning, $\delta_1>0$
(notice also that $\delta_2>0$ in \fig{fig:GU}),
the spin polarization is approximately equal to $1$ and no sign change occurs.
This can be understood by realizing that the exchange field mimics the 
effect of an external magnetic field only when $\delta_1 / |\delta_1| =- \delta_2 / |\delta_2|$,
i.e., when the detuning in each dot has different sign; cf.
Eqs.~(\ref{eq:exch}) and (\ref{eq:exch2}).
Consequently, one should expect that the sign change of spin polarization will occur
when $\delta_1 \lessgtr 0$ and $\delta_2 \gtrless 0$.
This is indeed what we observe in the fully interacting case,
as can be seen in \fig{fig:GU2D} calculated for $U_1=U_2$,
which shows the spin polarization as a function of the
double quantum dot levels $\e_1$ and $\e_2$.

Figure \ref{fig:GU2D} clearly demonstrates all the features expected on the basis of
analytical formulas presented in Sec. \ref{sec:P}. The spin polarization is very large
(approximately equal to $1$) and may change sign (reaching $\pol=-1$) as a function 
of either $\e_1$ or $\e_2$. However, this sign change occurs when the detunings
$\delta_1$ and $\delta_2$ have opposite signs. Moreover, this interesting behavior 
of the spin polarization occurs when the second dot is in the local moment regime, 
$-U<\e_2<0$, irrespective of the first dot's occupancy. In other words, for any 
$\e_2$ such that $-U<\e_2<0$ and $\e_2 \neq -U_2/2$, there exists such $\e_1$ that the
spin polarization changes sign and becomes $-1$. This sign change occurs at the 
level position where the linear conductance exhibits a minimum. The magnitude of 
the conductance is then of the order of that in the cotunneling regime.

\subsection{Finite temperature}

\begin{figure}[t]
\centering
\includegraphics[width=0.9\columnwidth]{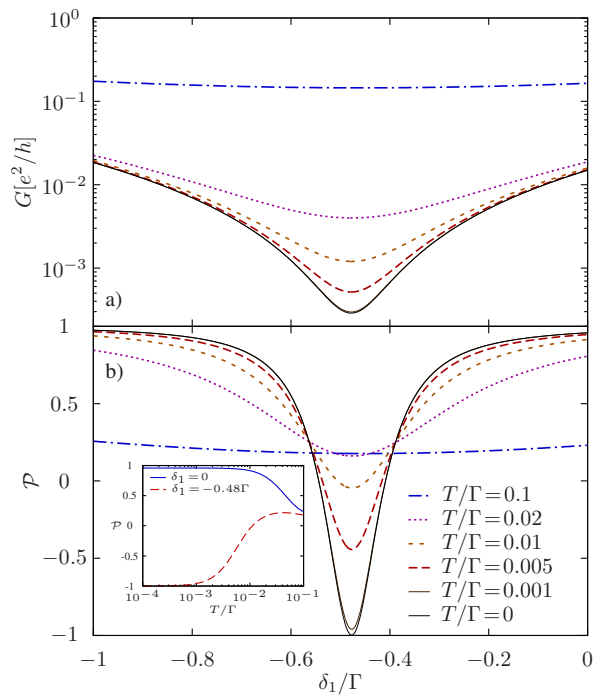}
\caption{(Color online)
		 The linear conductance (a) and the spin polarization (b)
         as a function of $\delta_1$ calculated for 
         different temperatures $T$ and for $U_1=U_2=U=0.5$. Inset in (b)
         presents the temperature dependence of $\pol$
         for $\delta_1 = 0$ and $\delta_1/\Gamma = -0.48$.
         The other parameters are the same as in \fig{fig:GU}.}
\label{fig:GT}
\end{figure}

Finally, we consider the effect of finite temperature on the operation of our 
spin-polarized current source. The $\delta_1$ dependence of the linear conductance
and the spin polarization calculated for different temperatures $T$ is shown in
\fig{fig:GT} for the fully interacting case with $U_1=U_2=U$. 
This figure was calculated for $\e_2=-U_2/3$, which implies that 
the exchange field is much larger than the Kondo temperature,
and the Kondo effect is suppressed. Thus, there is no universal energy scale.
Because the coupling $\G$ is directly measurable and determines another important 
energy scale, namely the exchange field, in \fig{fig:GT} we express 
the temperature in units of $\G=U/2$. One can see that by increasing $T$,
the conductance suppression becomes weakened, since thermal fluctuations generally
suppress the Fano effect. Consequently, the absolute value of the spin polarization
is also decreased. Moreover, the effect of the sign change of $\pol$, directly
associated with the spin-dependent Fano effect,
also becomes smeared out by finite temperature.
As can be seen in \fig{fig:GT}, the desired device operation persists only at low
temperatures, while already at $T=\G/10$, the conductance does not show any minimum
due to interference effects and the spin polarization is almost independent of 
$\delta_1$, with $\pol\approx p$.

The explicit dependence of the spin polarization for two representative level
detunings is shown in the inset of \fig{fig:GT}(b). For $\delta = -0.48\G$, 
$\pol = -1$ for $T\to 0$, however, once $T > \G/1000$, the absolute value of spin
polarization starts slowly decreasing. On the other hand, for $\delta=0$, the spin
polarization is equal to unity at low temperatures and decreases once $T > \G/100$. 
In fact, the relevant energy scale is given by the magnitude of the exchange field.
For realistic parameters, with $\G\sim$meV, the device should operate at clearly
cryogenic temperatures. However, for molecules, where both $U$ and $\Gamma$ can 
be larger, the relevant temperature range could be increased.

\section{Conclusions}
\label{sec:end}

In this paper, we have considered transport properties of T-shaped double quantum 
dots coupled to ferromagnetic leads. The calculations have been performed with the aid 
of the numerical renormalization group method, which allowed us to accurately
determine the spectral functions, the linear conductance, and the spin polarization 
of the current. Transport properties of the considered system are determined by the
Fano effect, which reveals itself as an antiresonance in linear conductance when
changing the DQD levels. On the other hand, the presence of ferromagnets results in
an exchange field that splits the levels in the dots. This results in
the spin dependence of the Fano effect -- the conditions for Fano destructive
interference are different in each spin channel. Because the
magnitude and sign of the exchange field can be controlled by changing the DQD's levels,
one can tune the conductance suppression in each spin channel. As a consequence,
there is a range of parameters where one of the conductances is much larger than 
the other one and the device exhibits perfect spin polarization. Moreover, because
the sign of the spin polarization can be changed by tuning the levels,
the operation of the device can be controlled by purely electrical means,
namely by appropriately sweeping the gate voltages. 
Our device thus provides a prospective example of an electrically controlled,
fully spin-polarized current source, which operates without the need to apply
external magnetic field.

From analytical analysis, we have found that to get perfect spin polarization, it 
is necessary to have finite Coulomb correlations in the dot, which is not directly
coupled to the leads (the second dot). Moreover, this dot should be in the local moment
regime, while no such restriction is imposed on the first dot, which can be
noninteracting. These findings have been confirmed by detailed NRG calculations,
which also revealed that finite Coulomb correlations in the first dot
can further increase the range of parameters where the sign change of
spin polarization occurs. Studying the conductance at finite temperatures,
we have shown that thermal fluctuations smear out the effects of interest,
which persist only at low temperatures.

Finally, we note that T-shaped DQDs can exhibit another interesting effects, such 
as, e.g., the two-stage Kondo effect.\cite{vojta02,cornagliaPRB05,zitkoPRB06,
chungPRB08} In this effect, with lowering temperature, at the first stage the spin 
in the first dot becomes screened by conduction electrons giving rise to maximum
conductance, and then, at lower temperatures, the second stage of screening occurs,
leading to conductance suppression. In fact, the
conductance suppression due to interference effects,
which occurs in T-shaped DQDs, can also be explained by invoking the two-stage 
Kondo effect \cite{sasakiPRL09,zitkoPRB10}.
However, a detailed analysis of the two-stage Kondo effect in the presence of
itinerant-electron ferromagnetism goes beyond the scope of the present paper and will
be considered elsewhere \cite{wojcik14}.


\acknowledgements

This work was supported by a 'Iuventus Plus' project No. IP2011 059471 in years
2012-2014 and the National Science Center
in Poland as the Project No. DEC-2013/10/E/ST3/00213.
I.W. also acknowledges support from the EU grant No. CIG-303 689.
Computing time at Pozna\'n Superconducting and Networking Center is acknowledged.


\end{document}